# Stokes flows in a sessile hemispherical drop due to evaporation and surface tension gradient

Peter Lebedev-Stepanov[1,α]

[1] Shubnikov Institute of Crystallography, Kurchatov Complex of Crystallography and Photonics of NRC'Kurchatov Institute', Leninskii prospekt 59, Moscow 119333, Russia

**Abstract** Viscous hydrodynamic flow in a small, slowly evaporating, sessile hemispherical droplet with a pinned contact line is considered. Analytical solutions are obtained for the Deegan outward flow, which is responsible for the "coffee ring effect", as well as the Marangoni flow excited by a surface tension gradient. It is assumed that the surface tension gradient may be caused by anisotropic cooling of droplet surface or other factors, such as nonuniform illumination of an optically active surfactant. Two main types of boundary conditions, no-slip and full-slip, are considered in describing the flow-substrate interaction. It is shown that under the no-slip condition, there is a rigid relationship between the evaporation rate and the surface tension gradient, which imposes strict requirements on the temperature regime inside the droplet. This result offers a new vision of the critical Marangoni number, which describes the threshold for the transition of an evaporating droplet from capillary flow to developed Marangoni convection. The results of this work may attract the attention of experimenters to the study of the sensitivity of viscous flow in an evaporating droplet to the liquid-substrate boundary conditions, especially if the system under consideration passes into the Marangoni regime, when the no-slip condition changes to a partial or full slip condition due to the increase in viscous shear stress near the substrate.

[α] e-mail: lebstep.p@crys.ras.ru (corresponding author)

# 1 Introduction

As many studies show, convective instability in a small evaporating sessile droplet is a complex and multifactorial phenomenon, the description of which cannot be reduced to simply calculating the Marangoni number and comparing it with some critical value [1]. The development of analytical approaches makes it possible to clarify the theoretical description of these processes by appealing to "first principles."

Much work has been devoted to studying the conditions under which the Marangoni convection occurs in sessile droplets. The importance of this topic is related to the numerous applications of such objects in inkjet printing [2], self-assembly technologies [3, 4], evaporative optical Marangoni assembly [5], evaporative lithography [6], thin-film coatings [7-8], medical diagnostics [9-11], etc. Modern numerical methods make it possible to study such systems over a wide range of parameters. However, the advantage of analytical models is ensured by their versatility, the possibility of flexible control of parameters and the order of approximation. Moreover, such first-principles models are fundamental in facilitating the search for new solutions to the fundamental equations of mathematical physics.

There are a number of special cases where analytical studies can be conducted using relatively simple mathematical constructions. First, this is the case of a flat droplet (small contact angles) that allows to apply the lubrication approximation. An analysis of this approximation, applicable to studies of Marangoni flows in evaporating droplets, is given in Ref. [12]. Another remarkable application of analytical methods is the possibility of using a spherical coordinate system, where the solution can be expressed using well-developed Legendre polynomials and other spherical functions.

A hemispherical droplet, whose shape is a compromise between the hydrophobic and hydrophilic type of substrate wetting, is a convenient object for studying flow structure depending on the nature of the flow-substrate interaction, the evaporation rate, and the surface tension gradient. Solutions obtained for a hemispherical droplet can form the basis for semiquantitative estimates of similar solutions for droplets with both smaller and larger contact angles than π/2. Therefore, analytical studies of hemispherical evaporating droplets have been conducted repeatedly (for example, in Ref. [13], the potential radial flow caused by evaporation is calculated).

This paper presents a study of the dependence of the structure of viscous Stokes flow (creeping flow) on the boundary conditions on the substrate and the hemispherical surface of the droplet. Typically, the greatest interest is evoked by the study of the toroidal Marangoni flow.

To determine the conditions for the occurrence of the Marangoni flow, it is necessary to consider the accompanying relatively weak flow, the so-called capillary flow [14], which leads to the "coffee ring effect." This type of flow was first described in the work of R. Deegan et al. [15] and is sometimes called the Deegan flow [16].

The Deegan flow arises as a result of two simultaneous processes: a decrease in droplet volume due to evaporation and the equalization of the droplet shape as a result of capillary, gravitational, and other forces applied to the sessile droplet. In other words, this flow ensures the spreading of an evaporating droplet in accordance with the behavior of its contact line. Given this, we believe it is more appropriate to use the term compensatory flow, meaning that the spreading corresponding to the contact line scenario compensates for the loss of volume caused by evaporation.

For droplets small compared to the capillary constant, capillary forces play the primary role in the formation of the Deegan flow. However, generally speaking, gravity plays a significant role in shaping the droplet, as do the adhesion forces between the liquid and the substrate, which are responsible for the behavior of the contact line. Two characteristic contact line scenarios are typically distinguished: CCR (constant contact radius, or contact line pinning) and CCA (constant contact angle) [17, 18].

Unlike the Deegan flow, the Marangoni flow is driven by a surface tension gradient,

which can be of any nature: thermal, solutal, photochemical, etc. [5, 12, 19]. Therefore, simply put, Marangoni flow occurs when boundary conditions associated with the surface tension gradient become dominant. It is generally accepted that Marangoni convection in a sessile droplet has a threshold nature and is described by a certain critical Marangoni number. Apparently, the quantitative value of this parameter depends on many parameters, which is a topic of debate.

To clarify the conditions for the occurrence of Marangoni flow, it is necessary to more accurately study its difference from capillary flow. In this paper, we show that under the no-slip boundary condition on the substrate, the Deegan flow corresponds to a certain surface tension gradient, which can be characterized by some Marangoni number. Thus, in this case, the Deegan and Marangoni flows are inseparable. Only by disabling the no-slip condition can these flows be separated. This conclusion prompts a new look at the threshold nature of Marangoni flow in an evaporating droplet.

The paper is structured as follows. Section 2 presents the problem statement, followed by a solution to the axisymmetric Stokes equations with no-slip boundary conditions in an evaporating hemispherical droplet with pinned contact line. The streamlines of the resulting flow, which combines the features of the Deegan flow with those of Marangoni flow, are visualized, and the properties of the solution are examined. The following sections examine the heat transfer problem, calculating the temperature and the corresponding surface tension gradient. The final section of the paper examines a system in which liquid sliding on a substrate is allowed, and analytical solutions for the both types of flow are obtained.

## 2. Problem Statement

Consider a small droplet deposited on a flat, solid, horizontal substrate. The equilibrium shape of a sessile droplet of a slowly evaporating liquid approximately corresponds to a spherical segment with a given contact (wetting) angle if the Bond number is $\text{Bo} = R^2 a^{-2} \ll 1$, where $a = (2\sigma)^{\frac{1}{2}}(\rho g)^{-\frac{1}{2}}$ is the capillary constant, and σ and ρ are the surface tension and mass density of the liquid, respectively; g is the acceleration due to gravity; and R is the droplet radius, respectively [18], Fig. 1.

The capillary constant for water is approximately 3.8 mm. If R does not exceed 1 mm for a given drop, then the condition is obviously satisfied. The origin of the spherical coordinate system is placed at the center of a sphere, the upper half of which represents the droplet under consideration. The substrate is a plane passing horizontally through the chosen center of the droplet. The z-axis is directed vertically upward from the center of the sphere, and the angle of deviation of the radius vector of length $r$ from the z-axis is the polar angle $\theta$ (Fig. 1).

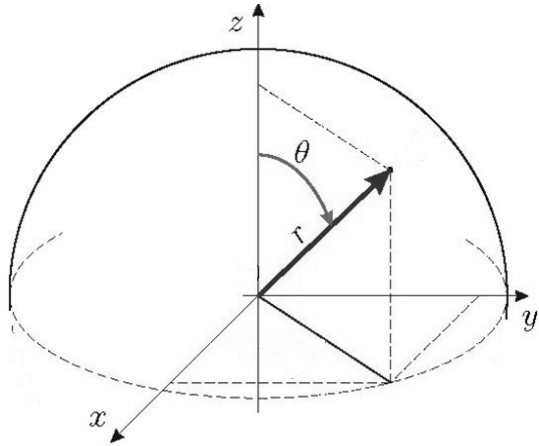

Fig.1 Geometry of a spherical coordinate system describing a droplet on a flat substrate.

The linearized axisymmetric stationary Navier–Stokes equations for an incompressible liquid (Stokes equations) in the spherical coordinate system have the form [20]:

$$\nabla^2 V_r - \frac{2 V_r}{r^2} - \frac{2}{r^2 \sin\theta} \frac{\partial}{\partial \theta}(V_\theta \sin\theta) = \eta^{-1} \frac{\partial}{\partial r} p, \qquad (1)$$

$$\nabla^2 V_\theta - \frac{V_\theta}{r^2 \sin^2\theta} + \frac{2}{r^2} \frac{\partial}{\partial \theta} V_r = \eta^{-1} \frac{1}{r} \frac{\partial}{\partial \theta} p, \qquad (2)$$

$$\frac{1}{r} \frac{\partial}{\partial r}(r^2 V_r) + \frac{1}{\sin\theta} \frac{\partial}{\partial \theta}(V_\theta \sin\theta) = 0, \qquad (3)$$

where $\eta$ is the dynamic viscosity, the Laplacian is

$$\nabla^2 = \frac{1}{r^2} \frac{\partial}{\partial r}\left(r^2 \frac{\partial}{\partial r}\right) + \frac{1}{r^2 \sin\theta} \frac{\partial}{\partial \theta}\left(\sin\theta \frac{\partial}{\partial \theta}\right), \qquad (5)$$

The velocity components $V_r(r,\theta)$ and $V_\theta(r,\theta)$, as well as the pressure $p(r,\theta)$, are independent of the azimuthal angle φ (Fig. 1). Here, formula (3) is an explicit representation of the continuity equation $(\nabla \cdot \mathbf{V}) = 0$.

The approximation corresponding to Eqs. (1)-(3) requires a small Reynolds number:

$$\text{Re} = \frac{RV\rho}{\eta} \ll 1, \tag{5}$$

where $V$ is the characteristic liquid flow velocity, $\rho$ is the liquid density, and $R$ is the droplet radius.

The boundary conditions on the substrate for a viscous liquid are:

$$V_r(r, \tfrac{\pi}{2}) = 0, \tag{6}$$

$$V_\theta(r, \tfrac{\pi}{2}) = 0. \tag{7}$$

The condition on a hemispherical surface, concerning the radial velocity, can generally be represented as an expansion in the Legendre polynomials

$$V_r(R, \theta) = \sum_{l=1}^{\infty} f_l P_l. \tag{8}$$

Eq. (8) describes the evolution of the droplet surface. If the droplet does not evaporate, the right-hand side is zero.

## 3 Derivation of the flow velocity

The general solution to the interior problem of equations (1)-(3) for the velocity components, stream function, and hydrodynamic pressure is [21]

$$V_r = \sum_{l=1}^{\infty} l(l+1) \left\{ \frac{a_l}{4l+6} \left(\frac{r}{R}\right)^{l+1} + c_l \left(\frac{r}{R}\right)^{l-1} \right\} P_l, \tag{9}$$

$$V_\theta = \sum_{l=1}^{\infty} \left\{ a_l \frac{l+3}{4l+6} \left(\frac{r}{R}\right)^{l+1} + c_l(l+1) \left(\frac{r}{R}\right)^{l-1} \right\} P_l^1, \tag{10}$$

$$\Psi(r, \theta) = -R^2 \sin\theta \sum_{l=1}^{\infty} \left\{ \frac{a_l}{4l+6} \left(\frac{r}{R}\right)^{l+3} + c_l \left(\frac{r}{R}\right)^{l+1} \right\} P_l^1(\cos\theta), \tag{11}$$

$$p(r, \theta) = \frac{\eta}{R} \sum_{l=0}^{\infty} (l+1) a_l \left(\frac{r}{R}\right)^l P_l(\cos\theta). \tag{12}$$

Using Eq. (9), condition (8) on a hemispherical surface can be rewritten as

$$V_r(R,\theta) = f(\theta) = \sum_{l=1}^{\infty} l(l+1)\left\{\frac{a_l}{4l+6} + c_l\right\} P_l = \sum_{l=1}^{\infty} f_l P_l. \tag{13}$$

From Eq. (13) one can obtain the relationship between the coefficients

$$\frac{a_l}{4l+6} + c_l = \frac{f_l}{l(l+1)}, \quad l = 1, 2, 3, \ldots. \tag{14}$$

The solution presented in the form given by Eqs. (9)-(10) is convenient for boundary conditions that fix the radial coordinate. However, if the angle is fixed, it is necessary to construct a series in powers of the radial coordinate:

$$V_r = \sum_{l=1}^{\infty} \left(\frac{r}{R}\right)^{l+1} \left\{\frac{a_l l(l+1)}{4l+6} P_l + (l+2)(l+3) c_{l+2} P_{l+2}\right\} + 2c_1 P_1 + 6c_2 \frac{r}{R} P_2, \tag{15}$$

$$V_\theta = \sum_{l=1}^{\infty} \left(\frac{r}{R}\right)^{l+1} \left\{a_l \frac{l+3}{4l+6} P_l^1 + c_{l+2}(l+3) P_{l+2}^1\right\} + 2c_1 P_1^1 + 3c_2 \frac{r}{R} P_2^1, \tag{16}$$

$$\Psi = -R^2 \sin\theta \sum_{l=1}^{\infty} \left(\frac{r}{R}\right)^{l+3} \left\{\frac{a_l}{4l+6} P_l^1 + c_{l+2} P_{l+2}^1\right\} - c_1 r^2 \sin\theta P_1^1 - c_2 \frac{r^3}{R} \sin\theta P_2^1. \tag{17}$$

Substituting (15) into condition (6), we obtain (Appendix 1)

$$V_r(r, \tfrac{\pi}{2}) = \sum_{n=1}^{\infty} \left(\frac{r}{R}\right)^{2n+1} \frac{(-1)^n (2n)!}{2^{2n} (n!)^2} (2n+1) \left\{\frac{a_{2n} n}{4n+3} - (2n+3) c_{2n+2}\right\} - 3c_2 \frac{r}{R} = 0 \tag{18}$$

This yields the condition for even coefficients

$$\frac{a_{2n} n}{4n+3} = (2n+3) c_{2n+2}, \quad n = 1, 2, 3, \ldots; \quad c_2 = 0. \tag{19}$$

Similarly, substituting Eq. (16) into condition (7), we obtain (Appendix 2):

$$V_\theta(r, \tfrac{\pi}{2}) = \sum_{n=1}^{\infty} \left(\frac{r}{R}\right)^{2n} \frac{(-1)^n (2n+2)(2n)!}{2^{2n} n!(n-1)!} \left\{\frac{a_{2n-1}}{4n+1} - c_{2n+1} \frac{2n+1}{n}\right\} - 2c_1 = 0. \tag{20}$$

This yields the condition for odd coefficients

$$\frac{a_{2n-1}}{4n+1} = c_{2n+1} \frac{2n+1}{n}, \quad n = 1, 2, 3, \ldots, \quad c_1 = 0. \tag{21}$$

Eqs. (19) and (21), together with Eq. (14), allow us to determine the desired coefficients $a_l$ and $c_l$ in term of $f_l$.

Transforming Eqs. (15)-(17) into sums containing even and odd polynomials, using relations (19) and (21), one can obtain relations for the velocity components expressed only in terms of the coefficients $a_l$:

$$V_r = \sum_{n=1}^{\infty} \left(\frac{r}{R}\right)^{2n+1} \frac{a_{2n} n}{4n+3}\{(2n+1)P_{2n} + (2n+2)P_{2n+2}\} + \sum_{n=1}^{\infty} \left(\frac{r}{R}\right)^{2n} \frac{a_{2n-1} n}{4n+1}\{(2n-1)P_{2n-1} + (2n+2)P_{2n+1}\} \quad (22)$$

$$V_\theta = \sum_{n=1}^{\infty} \left(\frac{r}{R}\right)^{2n+1} \frac{a_{2n}}{8n+6}\{(2n+3)P_{2n}^1 + 2nP_{2n+2}^1\} + \sum_{n=1}^{\infty} \left(\frac{r}{R}\right)^{2n} \frac{(n+1)a_{2n-1}}{4n+1}\left\{P_{2n-1}^1 + \frac{2n}{2n+1}P_{2n+1}^1\right\}. \quad (23)$$

$$\Psi = -R^2 \sin\theta \sum_{n=1}^{\infty} \left(\frac{r}{R}\right)^{2n+3} \frac{a_{2n}}{8n+6}\left\{P_{2n}^1 + \frac{2n}{(2n+3)}P_{2n+2}^1\right\} - R^2 \sin\theta \sum_{n=1}^{\infty} \left(\frac{r}{R}\right)^{2n+2} \frac{a_{2n-1}}{8n+2}\left\{P_{2n-1}^1 + \frac{2na_{2n-1}}{2n+1}P_{2n+1}^1\right\} \quad (24)$$

Using Eqs. (6) and (7), one can show that

$$\Psi(R, \tfrac{\pi}{2}) = 0. \quad (25)$$

Transforming expression (22) into a sum over polynomials:

$$V_r(r,\theta) = \left(\frac{r}{R}\right)^2 \frac{a_1}{5} P_1 + \left(\frac{r}{R}\right)^3 \frac{3a_2}{7} P_2 + \sum_{n=2}^{\infty}\left\{\left(\frac{r}{R}\right)^{2n+1} \frac{a_{2n} n(2n+1)}{4n+3} + \left(\frac{r}{R}\right)^{2n-1} \frac{a_{2n-2} 2n(n-1)}{4n-1}\right\} P_{2n} +$$
$$+ \sum_{n=2}^{\infty}\left\{\left(\frac{r}{R}\right)^{2n} \frac{a_{2n-1} n(2n-1)}{4n+1} + \left(\frac{r}{R}\right)^{2n-2} \frac{a_{2n-3} 2n(n-1)}{4n-3}\right\} P_{2n-1} \quad (26)$$

$$V_\theta = \left(\frac{r}{R}\right)^2 \frac{2a_1}{5} P_1^1 + \left(\frac{r}{R}\right)^3 \frac{5a_2}{14} P_2^1 + \sum_{n=2}^{\infty}\left\{\left(\frac{r}{R}\right)^{2n-1} \frac{na_{2n-2}}{4n-1} + \left(\frac{r}{R}\right)^{2n+1} \frac{(2n+3)a_{2n}}{8n+6}\right\} P_{2n}^1 +$$
$$+ \sum_{n=2}^{\infty}\left\{\left(\frac{r}{R}\right)^{2n} \frac{(n+1)a_{2n-1}}{4n+1} + \left(\frac{r}{R}\right)^{2n-2} \frac{2n(n-1)a_{2n-3}}{(4n-3)(2n-1)}\right\} P_{2n-1}^1 \quad (27)$$

$$\Psi = -R^2 \sin\theta \left(\frac{r}{R}\right)^4 \frac{a_1 P_1^1}{10} - R^2 \sin\theta \left(\frac{r}{R}\right)^5 \frac{a_2 P_2^1}{14} - R^2 \sin\theta \sum_{k=2}^{\infty}\left\{\left(\frac{r}{R}\right)^{2n+1} \frac{(n-1)a_{2n-2}}{(4n-1)(2n+1)} + \left(\frac{r}{R}\right)^{2n+3} \frac{a_{2n}}{8n+6}\right\} P_{2n}^1 -$$
$$- R^2 \sin\theta \sum_{n=2}^{\infty}\left\{\left(\frac{r}{R}\right)^{2n} \frac{(n-1)a_{2n-3}}{(4n-3)(2n-1)} + \left(\frac{r}{R}\right)^{2n+2} \frac{a_{2n-1}}{8n+2}\right\} P_{2n-1}^1 \quad (28)$$

At points on the droplet surface, Eq. (26) yields

$$V_r(R,\theta) = \frac{a_1}{5}P_1 + \frac{3a_2}{7}P_2 + \sum_{n=2}^{\infty}\left\{\frac{a_{2n}n(2n+1)}{4n+3} + \frac{a_{2n-2}2n(n-1)}{4n-1}\right\}P_{2n} +$$
$$+\sum_{n=2}^{\infty}\left\{\frac{a_{2n-1}n(2n-1)}{4n+1} + \frac{a_{2n-3}2n(n-1)}{4n-3}\right\}P_{2n-1} \tag{29}$$

Comparing (8) and (29), one can obtain

$$f_1 = \frac{a_1}{5}, \quad f_2 = \frac{3a_2}{7}, \tag{30}$$

$$f_{2n-1} = \frac{a_{2n-1}n(2n-1)}{4n+1} + \frac{a_{2n-3}2n(n-1)}{4n-3}, \quad n=2,3,4,\ldots \tag{31}$$

$$f_{2n} = \frac{a_{2n}n(2n+1)}{4n+3} + \frac{a_{2n-2}2n(n-1)}{4n-1}, \quad n=2,3,4,\ldots \tag{32}$$

Eqs.(30)-(32) demonstrate that if $f(\theta) = 0$ (the droplet does not evaporate and/or its surface is non-hemispherical), then only the trivial solution, $V_r(r,\theta) = V_\theta(r,\theta) = 0$, satisfies the no-slip boundary conditions (5), (6) and $V_r(R,\theta) = 0$.

## 4. Deegan outward flow

Let us consider the case where heat absorption on the droplet surface and the resulting temperature gradient can be neglected. Then the boundary condition on the droplet surface leads to the Deegan flow. The potential flow of an inviscid liquid in such a system was considered in the Ref. [13]. Here, we obtain a solution for Stokes flow that takes into account the droplet viscosity and the vortex velocity component.

Evaporation from a hemispherical surface in the isothermal approximation creates a flux density of the volatile component determined by the formula [18]

$$J_0 = \frac{Dn_S(1-\chi)}{R}, \tag{33}$$

where $D$ and $n_S$ are the diffusion coefficient and the volume concentration of saturated vapor in the air above the spherical surface of the droplet, respectively, and $\chi$ is the relative humidity of the air (if an aqueous droplet is considered).

The loss of liquid volume due to evaporation per unit time, taking into account Eq.

(33), is determined by the relationship:

$$W = 2\pi R^2 \int_0^{\frac{\pi}{2}} J_0 \sin\theta d\theta = 2\pi R D n_S (1-\chi). \qquad (34)$$

Let us consider the boundary conditions corresponding to a fixed contact line that does not shift during droplet evaporation. In this case, the droplet flattens as it evaporates. The volume (34) evaporated per unit time is redistributed, and the following expression must be satisfied:

$$W = 2\pi R^2 \int_0^{\frac{\pi}{2}} U(\theta) \sin\theta d\theta, \qquad (35)$$

where $U(\theta)$ is a smooth function that ensures that the boundary conditions for the liquid during evaporation are satisfied, and the liquid velocity on the surface is determined by the relationship:

$$V_r(R,\theta) = \frac{J(\theta)}{n_L} - U(\theta). \qquad (36)$$

where $\rho_L$ is the liquid density. Taking into account Eqs. (34) and (35), the following relation must be satisfied

$$\int_0^{\frac{\pi}{2}} V_r(R,\theta) \sin\theta d\theta = 0. \qquad (37)$$

A boundary condition, corresponding to contact line pinning, proposed in Ref. [13] for potential flow in a hemispherical droplet, takes the form

$$V_r(R,\theta) = \frac{2J_0}{n_L}\left(\frac{1}{2} - \cos\theta\right), \qquad (38)$$

Obviously, Eq. (38) satisfies condition (37). However, it does not satisfy condition (6), i.e., it is nonzero at a point on the substrate at $\theta = \frac{\pi}{2}$, and therefore is not applicable to describing a viscous liquid flowing without slip.

Let us find a boundary condition that satisfies the problem under consideration. For a hemispherical droplet, if the surface temperature is constant, the evaporation rate is independent of the polar angle θ. At a point on the substrate, $\theta = \frac{\pi}{2}$, evaporation is zero. Then, the evaporation flux density is described by a function of the form

$$J(\theta) = \begin{cases} J_0 = \text{const}, & 0 \leq \theta < \frac{\pi}{2} \\ 0, & \theta = \frac{\pi}{2} \end{cases} \quad (39)$$

It can be shown that a function satisfying conditions (39) can be represented as an expansion in odd Legendre polynomials (Appendix 3):

$$J(\theta) = J_0 \sum_{n=0}^{\infty} (-1)^n \frac{(4n+3)(2n)!}{(n+1)2^{2n+1}(n!)^2} P_{2n+1}(\cos\theta). \quad (40)$$

Then if

$$U(\theta) = \sum_{n=0}^{\infty} B_{2n+1} P_{2n+1}, \quad (41)$$

where $B_{2n+1}$ are constant coefficients, satisfies (44) for

$$W = 2\pi R D n_S (1-\chi), \quad (42)$$

then expression (36), composed of Eqs. (40) and (41), satisfies the boundary condition for evaporation (37) and condition (6).

To normalize (37), taking into account Eqs. (38) and (6), and also considering that the expression holds

$$\int_0^{\frac{\pi}{2}} \cos\theta \sin\theta d\theta = \frac{1}{2} \quad (43)$$

one has to accept

$$B_1 = -\frac{2J_0}{n_L}, \quad B_i = 0, \ i = 2,3,4,.... \quad (44)$$

The velocity on the droplet surface can be represented as

$$V_r(R,\theta) = \frac{J(\theta)}{n_L} - \frac{2J_0}{n_L}\cos\theta. \quad (45)$$

Separating the polynomial $P_1 = \cos\theta$ from Eq. (45), we obtain

$$V_r(R,\theta) = -\frac{u_0}{2}P_1 + u_0\sum_{n=1}^{\infty}(-1)^n \frac{(4n+3)(2n)!}{(n+1)2^{2n+1}(n!)^2}P_{2n+1}, \quad u_0 = \frac{J_0}{n_L}. \tag{46}$$

Hence, the coefficients of expression (29), corresponding to even Legendre polynomials $P_{2n}$, are equal to zero:

$$f_{2k} = 0, \ k = 1, 2,... \tag{47}$$

and, taking into account (32),

$$a_{2k} = 0, \ k = 1, 2,... \tag{48}$$

Taking into account (46), we have

$$f_1 = -\frac{u_0}{2}, \quad a_1 = -\frac{5u_0}{2}, \tag{49}$$

Comparing Eq. (13) with Eq. (46), one can obtain

$$f_{2k+1} = u_0(-1)^k \frac{(4k+3)(2k)!}{(k+1)2^{2k+1}(k!)^2}, \ k = 1, 2, 3,... \tag{50}$$

Then, according to (31), for the remaining odd coefficients we have

$$a_{2k+1} = u_0 \frac{(-1)^k(4k+3)(4k+5)(2k)!}{2^{2k+1}(2k+1)((k+1)!)^2} - \frac{2k(4k+5)a_{2k-1}}{(4k+1)(2k+1)}, \ k = 1, 2, 3,.... \tag{51}$$

Recurrence relation (51) allows us to sequentially find all the remaining odd coefficients. The calculation yields

$$a_3 = \frac{27}{16}u_0, \quad a_5 = -\frac{65}{48}u_0, \quad a_7 = \frac{595}{512}u_0,$$

$$a_9 = -\frac{1323}{1280}u_0, \quad a_{11} = \frac{1925}{2048}u_0, \quad a_{13} = -\frac{12441}{14336}u_0 \tag{52}$$

We see that the Deegan flow caused by droplet evaporation is described by series with odd alternating coefficients with decreasing absolute values. The visualization of the current lines is shown in Fig. 2. The calculation is performed using formula (24), where the summation is carried out up to and including the term with the coefficient $a_{21}$.

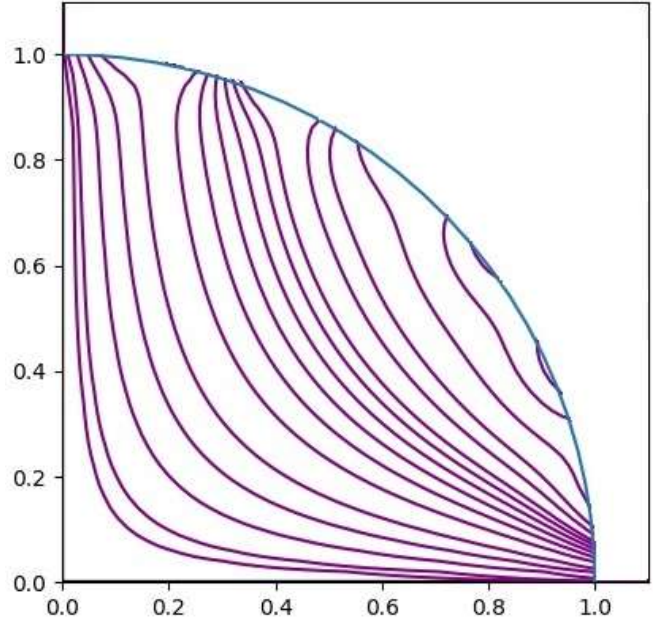

Fig.2. Streamlines of viscous flow in an evaporating droplet under no-slip conditions on a substrate

## 5. Regimes of the heat transfer

Heat transfer in a droplet occurs due to molecular thermal conductivity and liquid convection, which causes hydrodynamic mixing. Assuming that the droplet evaporates slowly, we will use the steady-state convective heat conduction equation [22]:

$$(\mathbf{V},\nabla)T = a\nabla^2 T, \qquad (53)$$

where $a = \dfrac{\lambda}{c_p \rho}$ is the thermal diffusivity coefficient ($\lambda$ is the thermal conductivity coefficient, $c_p$ is the specific heat capacity, and $\rho$ is the density of the liquid), and $\mathbf{V}$ is the liquid velocity.

Let the substrate have very high thermal conductivity and sufficient heat capacity so that lateral heat gradients can be neglected. In this case, the boundary condition on the substrate is

$$T(r, \tfrac{\pi}{2}) = T_0. \qquad (54)$$

Let us consider the heat flux on the hemispherical surface of the droplet. Due to liquid evaporation, the intensity of which, assuming a small temperature gradient on the surface, is determined by the function $J(\theta)$ describing by the formulae (42)-(44). The heat outflow per unit surface area is determined by the expression

$$q(\theta) = MHJ(\theta), \tag{55}$$

where $H$ is the specific enthalpy of evaporation, $M$ is the molecular mass of the liquid, so that mass density of liquid is $\rho = Mn_L$.

Since the thermal conductivity of air is orders of magnitude lower than that of the liquid, compensation for heat transfer (55) occurs primarily through the liquid from the substrate. This compensating heat input has two components: molecular and convective:

$$q(\theta) = -\lambda \frac{\partial T}{\partial r}\bigg|_{r=R} + c_p (T(R,\theta) - T_0)\rho V_r(R,\theta). \tag{56}$$

The Peclet number Pe defines the ratio of the intensities of convective and molecular heat transfer [22]

$$\text{Pe} = \frac{c_p \rho V R}{\lambda} = \frac{VR}{a}, \tag{57}$$

where $V$ is the characteristic velocity of the liquid.

If Pe $\ll$ 1 in equations (53) and (56), the terms responsible for convection can be neglected, so that heat transfer is described by the Laplace equation:

$$\nabla^2 T = 0, \tag{58}$$

where Laplace operator in the axisymmetric case in a spherical coordinate system has the form

$$\nabla^2 = \frac{1}{r^2}\frac{\partial}{\partial r}\left(r^2 \frac{\partial}{\partial r}\right) + \frac{1}{r^2 \sin\theta}\frac{\partial}{\partial \theta}\left(\sin\theta \frac{\partial}{\partial \theta}\right). \tag{59}$$

Boundary condition is

$$MHJ(\theta) = -\lambda \frac{\partial T}{\partial r}\bigg|_{r=R}. \tag{60}$$

The solution to equation (58) under boundary condition (54) is described by a sum of Legendre polynomials of odd degree, which vanish at $\theta = \frac{\pi}{2}$:

$$T(r,\theta) = T_0 + \sum_{n=0}^{\infty} F_{2n+1} \left(\frac{r}{R}\right)^{2n+1} P_{2n+1}(\cos\theta). \tag{61}$$

If Pe>>1, molecular thermal conductivity can be neglected, and heat transfer is determined by Marangoni convection. In this case, Eq. (53) takes the form:

$$(\mathbf{V}, \nabla)T = 0. \qquad (62)$$

Consider (62) in a spherical coordinate system for an axisymmetric system:

$$V_r \frac{\partial T}{\partial r} + \frac{V_\theta}{r} \frac{\partial T}{\partial \theta} = 0. \qquad (63)$$

The liquid velocity can be described by the stream function

$$V_r = \frac{1}{r^2 \sin\theta} \frac{\partial \Psi}{\partial \theta}, \quad V_\theta = -\frac{1}{r \sin\theta} \frac{\partial \Psi}{\partial r}. \qquad (64)$$

Substituting Eq. (64) into Eq. (63), we obtain

$$\frac{1}{r^2 \sin\theta} \frac{\partial \Psi}{\partial \theta} \frac{\partial T}{\partial r} - \frac{1}{r^2 \sin\theta} \frac{\partial \Psi}{\partial r} \frac{\partial T}{\partial \theta} = 0. \qquad (65)$$

Obviously, the solution to the Eq.(65) is

$$T(r, \theta) = A\Psi(r, \theta) + C, \qquad (66)$$

where $A$ and $C$ are the constants.

## 6. Characteristic temperature shift of the Deegan flow

The boundary condition on the free surface, neglecting the deviation of the droplet shape from spherical, has the form [23, 24]:

$$\left.\frac{\partial V_r}{\partial \theta}\right|_R + R \left.\frac{\partial V_\theta}{\partial r}\right|_R - V_\theta(R, \theta) = \frac{1}{\eta} \frac{\partial \sigma}{\partial \theta}. \qquad (67)$$

The surface tension of the liquid decreases with increasing temperature. The surface tension gradient along the surface can be written as

$$\frac{\partial \sigma(\theta)}{\partial \theta} = -\beta \frac{\partial T(R, \theta)}{\partial \theta} \quad \text{where} \quad \beta = -\frac{\partial \sigma}{\partial T}. \qquad (68)$$

Since the coefficients for the Deegan flow have already been found, it is interesting to consider the characteristic temperature shift corresponding to this flow, based on the condition (67). It should be kept in mind that, since the heat conduction problem is not being solved here, we are dealing with a certain conventional temperature drop characterizing the Deegan flow under consideration.

Considering that [25]

$$\frac{\partial P_k(\cos\theta)}{\partial \theta} = P_k^1(\cos\theta), \tag{69}$$

using (26) and (27), we find

$$\left.\frac{\partial V_r}{\partial \theta}\right|_R = \frac{a_1}{5}P_1^1 + \sum_{n=2}^{\infty}\left\{\frac{a_{2n-3} 2n(n-1)}{4n-3} + \frac{a_{2n-1}n(2n-1)}{4n+1}\right\}P_{2n-1}^1, \tag{70}$$

$$\left.R\frac{\partial V_\theta}{\partial r}\right|_R = \frac{4a_1}{5}P_1^1 + \sum_{n=2}^{\infty} 2n\left\{\frac{2(n-1)^2 a_{2n-3}}{(4n-3)(2n-1)} + \frac{(n+1)a_{2n-1}}{4n+1}\right\}P_{2n-1}^1, \tag{71}$$

$$V_\theta(R,\theta) = \frac{2a_1}{5}P_1^1 + \sum_{n=2}^{\infty}\left\{\frac{2n(n-1)a_{2n-3}}{(4n-3)(2n-1)} + \frac{(n+1)a_{2n-1}}{4n+1}\right\}P_{2n-1}^1. \tag{72}$$

Substituting Eq. (68) and Eqs. (70)-(72) into Eq. (67), we obtain

$$\frac{3a_1}{5}P_1^1 + \sum_{n=2}^{\infty}\left\{\frac{8n(n-1)^2}{(4n-3)(2n-1)}a_{2n-3} + \frac{(2n-1)(2n+1)}{4n+1}a_{2n-1}\right\}P_{2n-1}^1 = -\frac{\beta}{\eta}\frac{\partial T(R,\theta)}{\partial \theta}. \tag{73}$$

Integrating Eq. (73) over the angle $\theta$, we have

$$T(R,\theta) = -\frac{3\eta a_1}{5\beta}P_1 - \frac{\eta}{\beta}\sum_{n=2}^{\infty}\left\{\frac{8n(n-1)^2}{(4n-3)(2n-1)}a_{2n-3} + \frac{(2n-1)(2n+1)}{4n+1}a_{2n-1}\right\}P_{2n-1} + T_0. \tag{74}$$

Comparing Eq. (74) with Eq. (61), we obtain a solution of the form:

$$T(r,\theta) = -\frac{3\eta a_1}{5\beta}\frac{r}{R}P_1 - \frac{\eta}{\beta}\sum_{n=2}^{\infty}\left\{\frac{8n(n-1)^2}{(4n-3)(2n-1)}a_{2n-3} + \frac{(2n-1)(2n+1)}{4n+1}a_{2n-1}\right\}\left(\frac{r}{R}\right)^{2n-1}P_{2n-1} + T_0 \tag{75}$$

If a molecular heat conduction regime existed in a droplet at Pe << 1, leading to temperature field given by Eq. (75) due to the droplet cooling during evaporation, it would correspond to a hydrodynamic flow described by Eqs. (26)–(28), which are classified as capillary compensatory flow, similar to those introduced by Deegan et al [15]. This blurs the line between the Deegan and the Marangoni flows.

We emphasize that, generally speaking, Eq. (75) does not represent the actual temperature associated with the heat of evaporation of the solvent, but rather a certain conventional characteristic quasi-temperature calculated based on the velocity field of capillary flows (not Marangoni flow, as is the usual classification). This value can be estimated as:

$$\Delta T_D \propto \frac{\eta u_0}{\beta} = \frac{D\eta \rho_S (1-\chi)}{\beta \rho_L R}, \qquad (76)$$

where $T_0$ is the substrate temperature. In practically significant cases (a drop of water under normal conditions), this is a very small quantity in quantitative terms.

Quantity (76) provides an estimate of the temperature difference that causes a Marangoni flow equivalent to the compensation flow under consideration. In this sense, using Eq. (76), one can calculate the Marangoni number for a given Deegan flow.

The Marangoni number is determined by

$$\text{Ma} = \frac{c_p \rho_L \Delta T R}{\eta \lambda} \left| \frac{d\sigma}{dT} \right|. \qquad (77)$$

Substituting Eq. (76) into Eq. (77), we obtain the desired estimate of the effective Marangoni number for the Deegan flow

$$\text{Ma}_D = \frac{c_p \rho_S D (1-\chi)}{\lambda}. \qquad (78)$$

Thermodynamic and kinetic parameters of liquid water at normal conditions are taken or calculated from [3]: $c_p \approx 4.2 \text{ kJ} \cdot \text{kg}^{-1} \cdot \text{K}^{-1}$, $\eta \approx 10^{-3} \text{Pa} \cdot \text{c}$, $\lambda \approx 0.6 \text{ J} \cdot \text{m}^{-1} \text{c}^{-1} \text{K}^{-1}$, $\rho_S \approx 2 \cdot 10^{-2} \text{ kg} \cdot \text{m}^{-1}$, $\rho_L \approx 1000 \text{ kg} \cdot \text{m}^{-1}$, $D \approx 2.4 \cdot 10^{-5} \text{ m}^2 \cdot \text{c}^{-1}$, $\beta = \left| \frac{d\sigma}{dT} \right| \approx 1.5 \cdot 10^{-4} \text{ J} \cdot \text{m}^{-2} \text{K}^{-1}$, $\chi \approx 0.5$.

If the droplet has a radius of $R = 10^{-4} \text{m}$, then $\Delta T_D \approx 1.6 \cdot 10^{-5} \text{ K}$, and $Ma_D \approx 0.0017$. These are very small values that can usually be neglected compared to the typical values of temperature difference and Marangoni number characteristic of convective flow in an evaporating liquid droplet under normal conditions.

**7. Marangoni flow**

Let us now consider a solution in which the liquid is allowed to slide over the substrate, i.e., condition (6) is not satisfied, and therefore the requirement (19) that follows from it is not introduced. The substrate impermeability condition (7) and the corresponding condition for odd coefficients (21) remain valid. On the hemispherical surface of the droplet, boundary condition (38) is applicable in this case, which violates condition (6). In this case, $V_r(R,\theta)$ given by Eq. (38) can be represented as an expansion in even Legendre polynomials in the interval $\theta \in [0, \frac{\pi}{2}]$ proposed by Yu. Tarasevich in Ref. [13]:

$$G = \frac{1}{2} - \cos\theta = \sum_{n=1}^{\infty} (-1)^n \frac{(4n+1)(2n-2)!}{2^{2n}(n-1)!(n+1)!} P_{2n} = \sum_{n=1}^{\infty} G_n P_{2n}. \tag{79}$$

Then condition (38) takes the form

$$V_r(R,\theta) = \frac{2J_0}{n_L} \sum_{n=1}^{\infty} (-1)^n \frac{(4n+1)(2n-2)!}{2^{2n}(n-1)!(n+1)!} P_{2n}. \tag{80}$$

The odd coefficients of the expansion satisfy (21). Taking into account (29), they all vanish:

$$a_{2n+1} = c_{2n+1} = 0, \quad n = 0,1,2,\ldots. \tag{81}$$

On the other hand, according to Eq. (9), for even coefficients we have

$$V_r(R,\theta) = \sum_{k=1}^{\infty} 2k(2k+1)\left\{\frac{a_{2k}}{8k+6} + c_{2k}\right\} P_{2k}. \tag{82}$$

Equating the Eq. (80) and Eq.(82) yields

$$2k(2k+1)\left\{\frac{a_{2k}}{8k+6} + c_{2k}\right\} = \frac{2J_0}{n_L} G_k. \tag{83}$$

Using (9) and (10), we find

$$\left.\frac{\partial V_r}{\partial \theta}\right|_R = \sum_{l=1}^{\infty} l(l+1)\left\{\frac{a_l}{4l+6} + c_l\right\} P_l^1, \tag{84}$$

$$\left.R\frac{\partial V_\theta}{\partial r}\right|_R = a_1 \frac{4}{5} P_1^1 + \sum_{l=2}^{\infty} (l+1)\left\{a_l \frac{l+3}{4l+6} + c_l(l-1)\right\} P_l^1, \tag{85}$$

$$V_\theta(R,\theta) = \sum_{l=1}^{\infty} \left\{a_l \frac{l+3}{4l+6} + c_l(l+1)\right\} P_l^1. \tag{86}$$

Substituting Eqs. (68) and (84)-(86) into Eq. (67) and moving to even coefficients, we obtain

$$\sum_{l=1}^{\infty}\left\{a_l l\frac{l+2}{2l+3}+2c_l(l^2-1)\right\}P_l^1 = \sum_{k=1}^{\infty}\left\{a_{2k}k\frac{4k+4}{4k+3}+2c_{2k}(4k^2-1)\right\}P_{2k}^1 = -\frac{\beta}{\eta}\frac{\partial T(R,\theta)}{\partial\theta}. \quad (87)$$

According to Eq. (61), the main contribution to the temperature change can be represented as:

$$T(R,\theta)\approx T_0+A\cos\theta, \quad (88)$$

where A is a constant. Then, taking into account (79) and (69), Eq. (87) takes the form:

$$\frac{\partial T(R,\theta)}{\partial\theta}=-A\sin\theta=-A\sum_{n=1}^{\infty}(-1)^n\frac{(4n+1)(2n-2)!}{2^{2n}(n-1)!(n+1)!}P_{2n}^1=-A\sum_{n=1}^{\infty}G_n P_{2n}^1. \quad (89)$$

Substituting Eq. (89) into the right-hand side of (87), we obtain

$$a_{2k}k\frac{4k+4}{4k+3}+2c_{2k}(4k^2-1)=\frac{A\beta}{\eta}G_k. \quad (90)$$

Formulae (83) and (90) form a system of equations that allows us to calculate the coefficients:

$$a_{2k}=\left(\frac{A\beta}{2\eta}-\frac{J_0}{n_L}\frac{(2k-1)}{k}\right)\frac{(8k+6)G_k}{(4k+1)}, \quad (91)$$

$$c_{2k}=\left(\frac{J_0}{n_L}\frac{4(k+1)}{(2k+1)}-\frac{A\beta}{2\eta}\right)\frac{G_k}{4k+1}, \quad (92)$$

where according to Eq. (79)

$$G_k=(-1)^k\frac{(4k+1)(2k-2)!}{2^{2k}(k-1)!(k+1)!}. \quad (93)$$

The structure of coefficients (91)-(92) allows us to distinguish two velocity components: one responsible for the Deegan compensatory flow and one responsible for the Marangoni flow.

$$\mathbf{V}=\mathbf{V}_D+\mathbf{V}_M. \quad (94)$$

The expansion coefficients for $\mathbf{V}_D$ are as follows:

$$a_{D2k}=\frac{J_0}{n_L}(-1)^{k+1}\frac{(2k-1)(4k+3)(2k-2)!}{2^{2k-1}k(k-1)!(k+1)!}, \quad (95)$$

$$c_{D2k}=\frac{J_0}{n_L}\frac{(-1)^k(2k-2)!}{2^{2k-2}(2k+1)(k-1)!k!}, \quad (96)$$

and for $\mathbf{V}_M$ are determined by the formulas

$$a_{M2k} = \frac{A\beta}{\eta} \frac{(-1)^k (4k+3)(2k-2)!}{2^{2k}(k-1)!(k+1)!}, \tag{97}$$

$$c_{M2k} = -\frac{a_{M2k}}{8k+6}. \tag{98}$$

The stream function for the compensating flow according to Eq.(11) has the form:

$$\Psi_D(r,\theta) = -R^2 \sin\theta \sum_{k=1}^{\infty} \left\{ \frac{a_{D2k}}{8k+6} \left(\frac{r}{R}\right)^{2k+3} + c_{D2k} \left(\frac{r}{R}\right)^{2k+1} \right\} P_{2k}^1(\cos\theta). \tag{99}$$

Taking into account Eqs. (97)-(98), the stream function for the Marangoni flow can be represented as

$$\Psi_M(r,\theta) = R^2 \sin\theta \sum_{k=1}^{\infty} c_{M2k} \left(\frac{r}{R}\right)^{2k+1} \left\{ \left(\frac{r}{R}\right)^2 - 1 \right\} P_{2k}^1(\cos\theta). \tag{100}$$

In Fig. 3 (a) and (b), the streamlines given by Eqs. (99) and (100), respectively, are plotted in dimensionless format at $R = 1$.

Let us compare Fig. 2 and Fig. 3(a), which describe a capillary flow, but with different flow regimes near the substrate: in the first case, the no-slip condition defined by Eq. (6) is satisfied, while in the second case, this condition is not imposed, so we have a slip regime. We see that the streamlines in Fig. 2 are combined into bundles, so the flow is fractalized. This feature is not observed in Fig. 3(a), where we have a simpler picture. Obviously, if there is no surface tension gradient, i.e. $\beta = 0$, we will have a pure capillary flow in the absence of a surface tension gradient (i.e., what is shown in Fig. 3a). According to Eq. (77), it corresponds to $Ma = 0$.

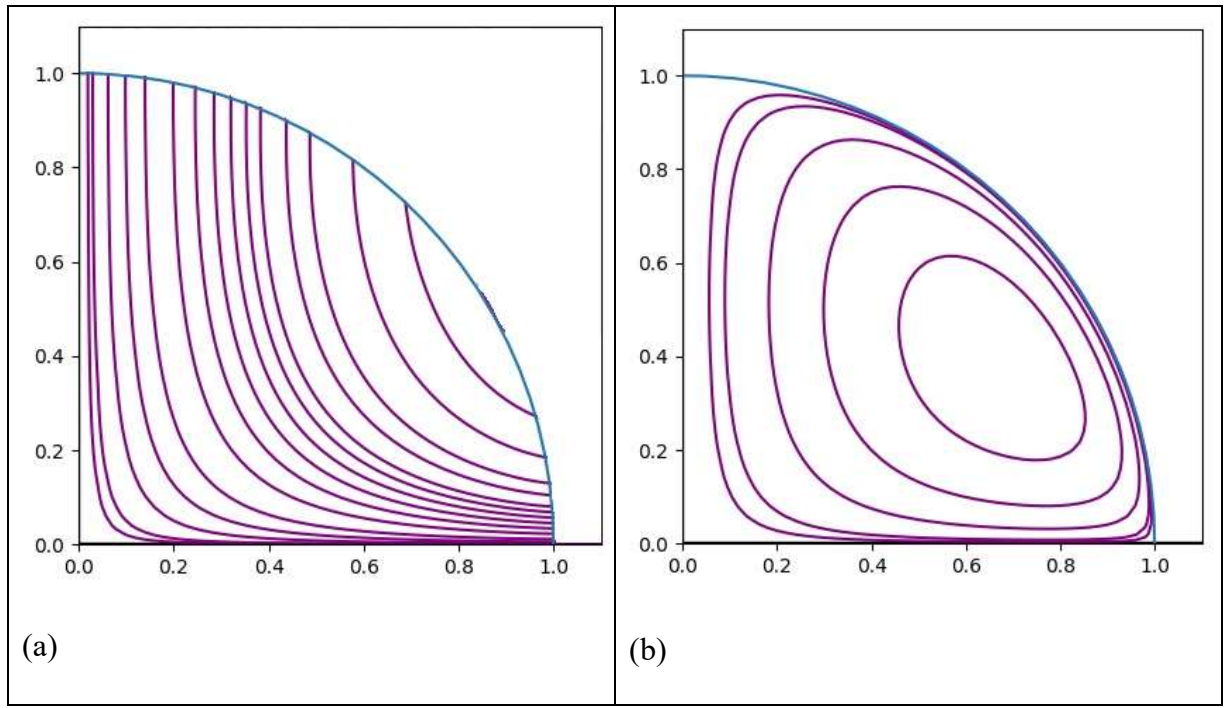

Fig.3. Линии тока двух типов вязкого течения в испаряющейся капле при условии slip на подложке: (a) – компенсационный поток, (b) – поток Марангони.

Thus, when using the slip condition on the substrate, the capillary flow is completely separated from the Marangoni flow and exhibits no singularities, whereas under the no-slip condition (6), the capillary flow has a non-zero Marangoni number (78) and therefore cannot be separated from the Marangoni flow. The no-slip condition creates a system strongly driven by evaporation, so introducing an independent parameter β becomes impossible.

Consider the case Pe<<1. Then, the constant A in Eq. (88) can be expressed from the first term of Eq. (61). Then, using Eq. (60), we obtain:

$$A = -\frac{3MHJ_0 R}{2\lambda}. \tag{101}$$

The physical meaning of Eq. (101) is the characteristic temperature difference between the droplet apex and the contact line (according to Eq. (88), the apex has a lower temperature than the substrate, which is maintained at a constant temperature $T_0$). Taking this into account, the Marangoni number (77) can be found using the formula

$$\mathrm{Ma} = \frac{MHJ_0 c_p \rho_L R^2 \beta}{\eta \lambda^2}, \tag{102}$$

and the Reynolds number (5) is

$$\mathrm{Re} = \frac{\rho MHJ_0 R^2 \beta}{\lambda \eta^2}. \tag{103}$$

Substituting the physical parameters, taking into account condition (5), leads to the conclusion that the approximation of purely molecular heat transfer, which is the basis for estimates (102) and (103), is suitable for describing the evaporation flows of viscous and poorly evaporating liquids such as oils, higher alcohols, and solvents such as ethylene glycol and glycerin.

For water droplets with radii of about 100 μm, the Peclet number can no longer be considered small, so convection dominates over molecular thermal conductivity. In this case, to estimate the constant A in formula (88) when solving the heat transfer equation (53), one can use the pure convection approximation (66).

## 8 Conclusion

An analytical solution is obtained for axisymmetric flows in a slowly evaporating sessile droplet with a pinned contact line. The study demonstrates the fundamental importance of the nature of the liquid-substrate adhesion: the observed flow essentially reflects this interaction.

The no-slip condition at the boundary (6) is either present or absent, so that the liquid is either stationary on the substrate or slides over it, respectively.

The main conclusion is that under the no-slip condition, there is a rigid relationship between the evaporation rate $J_0$ and the surface tension gradient β, which imposes strict requirements on the temperature regime inside the droplet. These requirements are not satisfied within the linear steady-state approximation and extend beyond it. Fig. 2 describes the structure of such a flow. It cannot be classified purely as a compensating flow, as it corresponds to a non-zero surface tension gradient and, consequently, a non-zero Marangoni number, described by Eq. (78). However, if we abandon the no-slip condition, evaporation rate $J_0$ and the tension gradient β become independent, so the flow can be separated into a compensatory flow (Fig. 3(a)) and a Marangoni flow (Fig. 3(b)), which are described independently.

The obtained result provides a new perspective on the critical Marangoni number, which describes the threshold for the transition of an evaporating droplet to developed Marangoni convection.

Obviously, in a real system, both no-slip and the slip, complete or partial, can occur. This is determined by the properties of the substrate and the liquid.

It is also possible that at low evaporation rates, i.e., a small temperature gradient and, correspondingly, low thermal loads, the no-slip condition holds, so the flow pattern is determined by Fig. 2. In this case, the Deegan flow and Marangoni flow are inseparable. However, if evaporation increases, the thermal load on the boundary conditions increases, which theoretically can cause the system to stall and transition to a sliding flow over the substrate. Then, a self-sustaining Marangoni flow and an independent classical Deegan flow (Fig. 3) emerge.

Obviously, the critical number Ma, corresponding to such a transition, has its own value for each specific system. For some systems, flow breakdown with low Ma can occur without breaking no-slip, but instead result in a transition to a transient self-oscillatory regime with a high Reynolds number, significant droplet surface deformation, and other nonlinear effects. The description of this regime goes beyond the assumptions underlying our study (the Stokes flow approximation and small droplet surface deformations).

The results of this work could encourage experimentalists to investigate the sensitivity of flow in an evaporating droplet to the liquid-substrate boundary condition, particularly when the system under consideration transitions from a regime with switched-off Marangoni flow to a regime where classical Marangoni flow is established (Fig. 3(b)), where this condition can change due to the increase in viscous shear stresses near the substrate.

**Data availability statement.** The data that support the findings of this study are available from the author upon reasonable request.

**Appendix 1**

Substituting Eq. (13) into condition (6), one can obtain

$$V_r(r, \tfrac{\pi}{2}) = \sum_{l=1}^{\infty} \left(\frac{r}{R}\right)^{l+1} \left\{ \frac{a_l l(l+1)}{4l+6} P_l(0) + (l+2)(l+3) c_{l+2} P_{l+2}(0) \right\} + 2c_1 P_1(0) + 6c_2 \frac{r}{R} P_2(0) = 0. \quad (1.1)$$

We have [25]:

$$P_{2n+1}(0) = 0, \quad (1.2)$$

$$P_{2n}(0) = \frac{(-1)^n (2n)!}{2^{2n} (n!)^2}. \quad (1.3)$$

Substituting Eqs. (1.2) and (1.3) in (1.1) and then sequentially transforming,

$$V_r(r, \tfrac{\pi}{2}) = \sum_{n=1}^{\infty} \left(\frac{r}{R}\right)^{2n+1} \left\{ \frac{a_{2n} 2n(2n+1)}{8n+6} P_{2n}(0) + (2n+2)(2n+3) c_{2n+2} P_{2n+2}(0) \right\} + 2c_1 P_1(0) + 6c_2 \frac{r}{R} P_2(0) = 0$$

$$V_r(r, \tfrac{\pi}{2}) = \sum_{n=1}^{\infty} \left(\frac{r}{R}\right)^{2n+1} \left\{ \frac{a_{2n} n(2n+1)}{4n+3} \frac{(-1)^n (2n)!}{2^{2n} (n!)^2} + (2n+2)(2n+3) c_{2n+2} \frac{(-1)^{n+1}(2n+2)!}{2^{2n+2}((n+1)!)^2} \right\} + 2c_1 P_1(0) + 6c_2 \frac{r}{R} P_2(0) = 0$$

$$V_r(r, \tfrac{\pi}{2}) = \sum_{n=1}^{\infty} \left(\frac{r}{R}\right)^{2n+1} \frac{(-1)^n (2n)!}{2^{2n}(n!)^2} \left\{ \frac{a_{2n} n(2n+1)}{4n+3} + (2n+2)(2n+3) c_{2n+2} \frac{(-1)(2n+1)(2n+2)}{2^2(n+1)^2} \right\} + 6c_2 \frac{r}{R} \frac{(-1)(2)!}{2^2(1!)^2} = 0$$

$$V_r(r, \tfrac{\pi}{2}) = \sum_{n=1}^{\infty} \left(\frac{r}{R}\right)^{2n+1} \frac{(-1)^n (2n)!}{2^{2n}(n!)^2} \left\{ \frac{a_{2n} n(2n+1)}{4n+3} - (2n+2)(2n+3) c_{2n+2} \frac{(2n+1)}{2(n+1)} \right\} - 3c_2 \frac{r}{R} = 0,$$

we obtain

$$V_r(r, \tfrac{\pi}{2}) = \sum_{n=1}^{\infty} \left(\frac{r}{R}\right)^{2n+1} \frac{(-1)^n (2n)!}{2^{2n}(n!)^2} (2n+1) \left\{ \frac{a_{2n} n}{4n+3} - (2n+3) c_{2n+2} \right\} - 3c_2 \frac{r}{R} = 0. \quad (1.4)$$

## Appendix 2

Substituting Eq. (14) into condition (7), one can obtain

$$V_\theta(r, \tfrac{\pi}{2}) = \sum_{l=1}^{\infty} \left(\frac{r}{R}\right)^{l+1} \left\{ a_l \frac{l+3}{4l+6} P_l^1(0) + c_{l+2}(l+3) P_{l+2}^1(0) \right\} + 2c_1 P_1^1(0) + 3c_2 \frac{r}{R} P_2^1(0) = 0, \quad (2.1)$$

where [25]

$$P_{2n}^1(0) = 0, \quad (2.2)$$

$$P_{2n-1}^1(0) = (-1)^n \frac{(2n)!}{2^{2n-1} n!(n-1)!}, \quad n = 1, 2, 3, \ldots. \quad (2.3)$$

Hence

$$V_\theta(r, \tfrac{\pi}{2}) = \sum_{n=1}^{\infty} \left(\frac{r}{R}\right)^{2n} \left\{ a_{2n-1} \frac{2n+2}{8n+2} P_{2n-1}^1(0) + c_{2n+1}(2n+2) P_{2n+1}^1(0) \right\} - 2c_1 = 0 \quad (2.4)$$

After sequentially transforming

$$V_\theta(r, \tfrac{\pi}{2}) = \sum_{n=1}^{\infty} \left(\frac{r}{R}\right)^{2n} (2n+2) \left\{ \frac{a_{2n-1}}{8n+2}(-1)^n \frac{(2n)!}{2^{2n-1} n!(n-1)!} + c_{2n+1}(-1)^{n+1} \frac{(2(n+1))!}{2^{2(n+1)-1}(n+1)! n!} \right\} - 2c_1 = 0$$

$$V_\theta(r, \tfrac{\pi}{2}) = \sum_{n=1}^{\infty} \left(\frac{r}{R}\right)^{2n} (-1)^n (2n+2) \frac{(2n)!}{2^{2n-1} n!(n-1)!} \left\{ \frac{a_{2n-1}}{8n+2} - c_{2n+1} \frac{(2n+1)(2n+2)}{2^2 n(n+1)} \right\} - 2c_1 = 0.$$

we obtain

$$V_\theta(r, \tfrac{\pi}{2}) = \sum_{n=1}^{\infty} \left(\frac{r}{R}\right)^{2n} \frac{(-1)^n (2n+2)(2n)!}{2^{2n} n!(n-1)!} \left\{ \frac{a_{2n-1}}{4n+1} - c_{2n+1} \frac{2n+1}{n} \right\} - 2c_1 = 0. \quad (2.5)$$

**Appendix 3**

Consider the auxiliary function

$$F(t) = \begin{cases} -1, & -1 \leq t < 0 \\ 0, & t = 0 \\ 1, & 0 < t \leq 1 \end{cases} \qquad (3.1)$$

The series expansion in Legendre polynomials for function (3.1) has the form

$$F(t) = \sum_{k=1}^{\infty} A_k P_k(t), \qquad (3.2)$$

where

$$A_k = \frac{2k+1}{2} \int_{-1}^{1} F(t) P_k(t) dt. \qquad (3.3)$$

Since the relation holds for Legendre polynomials $P_k(-t) = (-1)^k P_k(t)$, then for the antisymmetric function $F(t)$, only the coefficients (3.3) containing the Legendre polynomial with an odd index in the integral will be nonzero, so that:

$$A_{2n+1} = (2(2n+1)+1) \int_0^1 P_{2n+1}(t) dt, \quad a_{2n} = 0, \quad n=0,1,2,3,\ldots \qquad (3.4)$$

Let us use the differential equation for Legendre polynomials [25]

$$\frac{d}{dt}\left[(1-t^2)\frac{d}{dt}P_k(t)\right] = -k(k+1)P_k(t), \qquad (3.5)$$

$$A_k = -\frac{2k+1}{k(k+1)}(1-t^2)\frac{d}{dt}P_k(t)\bigg|_0^1 = \frac{2k+1}{k(k+1)}\frac{d}{dt}P_k(t)\bigg|_{t=0}. \qquad (3.6)$$

For further transformation, we use the recurrence formula [25]:

$$(1-t^2)\frac{d}{dt}P_k(t) = kP_{k-1}(t) - ktP_k(t). \qquad (3.7)$$

Hence

$$\frac{d}{dt}P_k(t)\bigg|_{t=0} = kP_{k-1}(0). \qquad (3.8)$$

Then

$$a_k = \frac{2k+1}{k+1} P_{k-1}(0). \tag{3.9}$$

For odd k=2n+1

$$A_{2n+1} = \frac{2(2n+1)+1}{2n+2} P_{2n}(0), \quad n=0,1,2,..., \tag{3.10}$$

where [25]

$$P_{2n}(0) = (-1)^n \frac{(2n)!}{2^{2n}(n!)^2}, \quad n=0,1,2,... \tag{3.11}$$

Then

$$A_{2n+1} = (-1)^n \frac{(4n+3)(2n)!}{(n+1)2^{2n+1}(n!)^2}, \quad n=0,1,2,... \tag{3.12}$$

Substituting (3.12) into (3.2), taking into account $A_{2n}=0$, we have

$$F(t) = \sum_{n=0}^{\infty} (-1)^n \frac{(4n+3)(2n)!}{(n+1)2^{2n+1}(n!)^2} P_{2n+1}(t), \tag{3.13}$$

Comparing (3.2) and (3.1), we finally find

$$J_0(\theta) = C \sum_{n=0}^{\infty} (-1)^n \frac{(4n+3)(2n)!}{(n+1)2^{2n+1}(n!)^2} P_{2n+1}(\cos\theta). \tag{3.14}$$